# Boundary curvature guided shape-programming kirigami sheets

Yaoye Hong[1], Yinding Chi[1], Yanbin Li[1], Yong Zhu[1], Jie Yin[1,*]

[a]Department of Mechanical and Aerospace Engineering, North Carolina State University, Raleigh, NC 27695, USA.

[*]To whom correspondence may be addressed. Email: jyin8@ncsu.edu

## Abstract

Kirigami, an ancient paper cutting art, offers a promising strategy for 2D-to-3D shape morphing through cut-guided deformation. Existing kirigami designs for target 3D curved shapes rely on intricate cut patterns in thin sheets, making the inverse design challenging. Motivated by the Gauss-Bonnet theorem that correlates the geodesic curvature along the boundary with the topological Gaussian curvature, here, we exploit programming the curvature of cut boundaries rather than complex cut patterns in kirigami sheets for target 3D curved topologies through both forward and inverse designs. Such a new strategy largely simplifies the inverse design. We demonstrate the achievement of varieties of dynamic 3D shape shifting under both mechanical stretching and remote magnetic actuation, and its potential application as an untethered predator-like kirigami soft robot. This study opens a new avenue to encode boundary curvatures for shape-programing materials with potential applications in shape-morphing structures, soft robots, and multifunctional devices.

## INTRODUCTION

Designing shape-programming materials from 2D thin sheets to 3D shapes has attracted broad and increasing interest in the past decades due to their novel materials properties imparted by geometrical shapes (*1*). Programmable shape shifting in various non-active and stimuli-responsive materials was realized at all scales utilizing folding, bending, and buckling (*2*). These shape-programmable materials are attractive for broad applications in programmable machines and robots (*3, 4*), functional biomedical devices (*5*), and four-dimensional (4D) printing (*6, 7*).

Kirigami, the ancient art of paper cutting, has recently emerged as a new promising approach for creating shape morphing structures and materials (*8, 9*). Cuts divide the original continuous thin sheets into discretized cut units but without sacrificing the structural integrity. Compared to continuous thin sheets, kirigami sheet enables more freedom and flexibility in shape shifting through local or global deformation between cut units (*10*). Starting from a thin sheet with patterned cuts, it can morph into varieties of 3D pop-up structures and surface topologies (*8, 9*) via rigid rotation mechanism (*11*) and/or out-of-plane buckling (*12*). The cuts impart new properties such as auxeticity (*13, 14*), stretchability (*15-20*), conformability (*16*), multistability (*21*), and optical chirality (*22*), which have found broad applications in mechanical metamaterials (*14, 20, 23*), stretchable devices (*16-18, 24, 25*), 3D mechanical self-assembly (*26*), tunable adhesion (*27*), and soft machines (*10, 28, 29*). The actuation of 2D-to-3D shape shifting can be achieved in both non-active and active materials in response to mechanical strains (*13, 14, 16*), environmental temperature (*10, 30, 31*), light (*29*), and magnetic field (*17, 20*).

Specifically, starting from a kirigami sheet or shell, 3D shapes with intrinsic curvature can be generated by utilizing non-uniform patterning and tessellation of polygon cut units through mechanical strains (*32, 33*) or pneumatic pressurization (*34*). The local heterogeneous deformation among non-periodic tessellated cut units induces global out-of-plane buckling of the 2D kirigami sheets, thus, resulting in the formation of different 3D curved shapes (*32, 33*). However, it often requires programming intricate cut patterns and arrangements of non-periodic cut units, making the inverse design and optimization for target 3D shapes complicated and challenging (*33, 34*).

Theoretically, the curvature of a boundary can be harnessed to tune 3D curved shapes based on the Gauss-Bonnet theorem in differential geometry (*35*), which correlates the topological Gaussian curvature and the geodesic curvature along the boundary (i.e., the projection of boundary curvature). Motivated by this theorem, here, we propose a new and simple strategy of utilizing the boundary curvature of cut edges rather than complex cut patterns to program 3D curved shapes through both forward and inverse designs. Unlike previous networked polygon cut units (*14, 16, 32-34*), our kirigami sheet is composed of parallel discrete ribbons enclosed by continuous boundaries (**Fig. 1 A**) through simple patterning of parallel cuts. We demonstrated that simply stretching the kirigami sheet with prescribed curved cut boundaries could generate varieties of well-predicted 3D curved shapes with positive, negative, and zero Gaussian curvatures and their combinations. The formed 3D shapes could be further reconfigured into other distinct configurations via bistability. We proposed a straightforward inverse design strategy for target 3D curved shapes by utilizing the geodesic feature of discrete ribbons. Finally, we demonstrated remote magnetic actuation to expand the shape-morphing capability and its proof-of-concept application in untethered soft robotics.

## RESULTS

### Manipulating 2D boundary curvatures for 3D curved topologies

The classical Gauss-Bonnet theorem (*35*) correlates the boundary curvature with the global topological Gaussian curvature $K$. Motivated by the theorem, as shown in **Fig. 1**, we start by designing the 2D precursors of kirigami sheets with different boundary curvatures $k_{bo}$ to exploit its effects on the Gaussian curvature of their 3D deployed shapes, where $k_{bo}$ is set to be positive (circular boundary in **Fig. 1 A**, ***i***), zero (rectangular boundary in **Fig. 1 B**, ***i***), and negative (biconcave circular boundary in **Fig. 1 C**, ***i***), respectively. We use the polyethylene terephthalate (PET) sheet with Young's modulus of 3.5 GPa, Poisson's ratio of 0.38, and thickness of 127 μm to fabricate the kirigami sheets using laser cutting (Supplementary Materials, Section S1). The thin sheets are cut into a number of discrete parallel thin ribbons enclosed by continuous boundary ribbon (ribbon width of 1.5 mm in **Fig. 1 A** and **B** and 0.75 mm in **Fig. 1 C**).

**Fig. 1 A, ii - C, ii** show that stretching the 2D precursors leads to distinct spheroidal, cylindrical, and saddle shapes with positive, zero, and negative Gaussian curvature $K$, respectively. Upon stretching, the boundary ribbon starts bending and compresses the enclosed discrete ribbons to induce their out-of-plane buckling. Thus, it renders a 3D pop-up topology. It should note that distinct from kirigami sheets composed of networked polygon cut units in previous studies (*14, 16,*

*32-34*), the simple design of parallel cuts in this work endows several unique characteristics: (1) It allows the sheet to be stretched along the direction perpendicular to the cuts via buckling of discrete ribbons. (2) Each bended discrete ribbon is bistable so that it could be tuned to either pop up or pop down for potential reconfiguration locally or globally (Fig. S1). (3) Parallel cuts make each discrete ribbon a geodesic curve of the morphed topology (Supplementary Materials, Section S2), facilitating inverse design as discussed later.

The observed 3D curved shapes can be qualitatively explained by the Gauss-Bonnet theorem. Mathematically, for the morphed 3D pop-up topology, the theorem can be simplified as

$$\int_\Omega K dA + \int_{\partial\Omega} k_{gb} ds = C \tag{1}$$

where the constant $C$ remains unchanged during shape shifting, and $k_{gb} = k_b \sin\varphi$ is the geodesic curvature along the boundary ribbon (*35*), i.e., the projection of the deformed boundary curvature $k_b$ with $\varphi$ being the projection angle (Supplementary Materials, Section S2). Thus, for 2D kirigami precursors with positive boundary curvature, i.e., $k_{bo} > 0$, as the applied strain increases, $k_{gb}$ decreases, resulting in a positive Gaussian curvature according to Eq. (1), i.e., $K > 0$ (see details in Supplementary Materials, Section S2), which is consistent with the observed spheroidal shape in **Fig. 1 *A, ii***. For the case of $k_{bo} = 0$, during the deformation, $k_{gb}$ remains zero, leading to a cylindrical shape with $K = 0$ as shown in **Fig. 1 *B, ii***. Similarly, for the case of $k_{bo} < 0$, $k_{gb}$ increases with the increasing applied strain. Thus, we have $K < 0$ in the observed saddle shape (**Fig. 1 *C, ii***).

**Analytical Modeling and Simulation on 3D Shape Shifting**

To quantify shape shifting of the kirigami structures with the applied strain, we combine both analytical modeling and finite element method (FEM) simulation to predict their topology changes (see details in Supplementary Materials, Section S1). The deformation of the kirigami structures is dominated by bending of the discrete ribbons, where the elastic strain energy in the boundary ribbon is negligible due to its small width (Supplementary Materials, Section S3). Thus, all the discrete ribbons share similar deformed elastica shapes (*36, 37*). The deformed 3D shape at an applied strain $\varepsilon$ can be described by $\vec{r}_s(\bar{s}_b, \bar{s}_d) = (\bar{x}(\bar{s}_b, \bar{s}_d), \bar{y}(\bar{s}_b, \bar{s}_d), \bar{z}(\bar{s}_b, \bar{s}_d))$, where $\bar{s}_b$ and $\bar{s}_d$ denote the normalized arc length coordinate of the boundary and the discrete ribbon as illustrated in **Fig. 1 A, i**, respectively. $(\overline{x}, \overline{y}, \overline{z})$ denote the Cartesian coordinates of any point $P(\overline{s}_b, \overline{s}_d)$ on the surface with its origin set at the center of the 2D precursor. Considering the deformed surface shape foliated by continuously varying discrete ribbons along the boundary, its generalized shape functions can be expressed as (see details in [Supplementary Materials, Section S3](#))

$$\overline{x}\left(\overline{s}_b, \overline{s}_d\right) = \frac{2m}{\lambda} CN\left(\lambda \overline{s}_d, m\right) \cos\alpha_1 + f\left(\overline{s}_b, \varepsilon\right) \qquad [2(a)]$$

$$\overline{y}\left(\overline{s}_b, \overline{s}_d\right) = \frac{2}{\lambda} E\left(AM\left(\lambda \overline{s}_d, m\right), m\right) - \overline{s}_d \qquad [2(b)]$$

$$\overline{z}\left(\overline{s}_b, \overline{s}_d\right) = \frac{2m}{\lambda} CN\left(\lambda \overline{s}_d, m\right) \sin\alpha_1 \qquad [2(c)]$$

by sweeping the varying discrete ribbons modeled as an elastica shape along the boundary. $m = m\left(\overline{s}_b, \varepsilon\right)$ is the elliptical modulus that characterizes the bending deformation of a discrete ribbon. $\lambda = 2F\left(\frac{\pi}{2}, m\right) / \overline{l}_d$ is related to the normalized length $\overline{l}_d$ of the discrete ribbon. $AM$ and $CN$ denote the Jacobian amplitude and the elliptic cosine, respectively. $E$ and $F$ denote the incomplete elliptic integral of the second kind and the first kind, respectively. $\alpha_1 = \alpha_1\left(\overline{s}_b, \varepsilon\right)$ is the tilting angle of the discrete ribbon with respect to the horizontal plane (i.e., $xy$ plane), which varies from 0 to 180º depending on its boundary location and the applied strain (**Fig. 1 A, ii**). $f\left(\overline{s}_b, \varepsilon\right)$ describes the $x$ coordinate at $\overline{s}_b$ of the deformed boundary ribbon.

Without losing generality, we can use three profiles from the front view, top view, and side view to characterize the 3D shape shifting with the applied strain (**Fig. 2 A - B**). The front view shows the backbone profile on the $xz$ plane (**Fig. 2 A, i - B, i**), which can be predicted by $\overline{x}_{bb} = \frac{2m}{\lambda} \cos\alpha_1 + f\left(\overline{s}_b, \varepsilon\right)$ and $\overline{z}_{bb} = \frac{2m}{\lambda} \sin\alpha_1$ after setting $\overline{s}_d = 0$ and $\overline{y} = 0$ in Eqs. 2. The top-view profile shows the deformed shape of the boundary ribbon (**Fig. 2 A, ii - B, ii**) that remains in the $xy$ plane during deformation by setting $\overline{z} = 0$ in Eqs. 2, which can be parametrized by

$$\tilde{r}_b(\overline{s}_b, \varepsilon) = (\overline{x}, \overline{y}, 0) = (f(\overline{s}_b, \varepsilon), g(\overline{s}_b, \varepsilon), 0) \qquad [3(a)]$$

where

$$g\left(\overline{s}_b, \varepsilon\right) = \left[\frac{2E\left(\frac{\pi}{2}, m\right)}{F\left(\frac{\pi}{2}, m\right)} - 1\right] g\left(\overline{s}_b, 0\right) \qquad [3(b)]$$

describes the $y$ coordinate at $\bar{s}_b$ of the deformed boundary ribbon at the strain of $\varepsilon$. Eq. 3(b) describes the relationship between $m$ and $\varepsilon$. Thus, combining Eqs. 2 and Eq. 3(b) will determine the unknown parameters of $\bar{x}$, $\bar{y}$, $\bar{z}$, and $m$ to predict the deformed 3D shapes with the applied strain. The side view shows the projection of similar elastica shapes of discrete ribbons onto the $yz$ plane (**Fig. 2 _A, iii_ - _B, iii_**), which depends on $m$ and the tilting angle of the longest discrete ribbon. Its deformed elastica shape can be expressed by $\bar{y}_d = \frac{2}{\lambda} E(AM(\lambda\bar{s}_d, m), m) - \bar{s}_d$ and $\bar{z}_d = \frac{2m}{\lambda} CN(\lambda\bar{s}_d, m)$, where the length of the discrete ribbons is assumed to be unchanged during deformation.

Next, we apply both the generalized analytical model and FEM simulation to analyze the 3D shape shifting in the specific examples shown in **Fig. 1**. **Fig. 2 _A, i-iii_** theoretically predict the variation of the three profiled shapes with the applied strain $\varepsilon$ during the formation of a spheroidal shape. As $\varepsilon$ increases from 0 to 0.4, top-view profiles show that the circular boundary gradually deforms into an elliptical shape (**Fig. 2 _A, ii_**), where we have $f(\bar{s}_b, \varepsilon) = (1 - \bar{w})\sin\bar{s}_b + \bar{v}\cos\bar{s}_b$ and $g(\bar{s}_b, \varepsilon) = (1 - \bar{w})\cos\bar{s}_b - \bar{v}\sin\bar{s}_b$ ($\bar{s}_b \in [-\frac{\pi}{2}, \frac{\pi}{2}]$) in the model (Supplementary Materials, Section S3). $\bar{w}(\bar{s}_b, \varepsilon)$ and $\bar{v}(\bar{s}_b, \varepsilon)$ denote the radial and tangential displacement of the boundary ribbon (*38*), respectively. Correspondingly, the compressed discrete ribbons deform into an elastica shape (side view in **Fig. 2 _A, iii_**). The backbone expands and shows an elliptical profile (front view in **Fig. 2 _A, i_**, Supplementary Materials, Section S2). As shown in **Fig. 2 _A, i-iii_**, the superposition of the three theoretically predicted front-view, top-view, and side-view profiles (highlighted in purple color) with images retrieved from the experimental observation at $\varepsilon = 0.3$ shows an excellent agreement. Similarly, the corresponding FEM simulated deformed 3D shape shows an excellent overlapping with the experiment (**Fig. 2 _A, iv_**).

**Fig. 2 _B, i-iii_** show the predicted shape change during the formation of a saddle shape. In contrast to simultaneous buckling in generating the spheroidal and cylindrical shapes (Supplementary Video S1-S2 and Fig. S6), we observe a sequential buckling during the formation of the saddle shape in experiments. The discrete ribbons near two stretching ends pop up first, followed by the ribbons in the center when beyond a critical strain $\varepsilon_c$ (Fig. S7). Such a sequential shape shifting is well captured by both the analytical model and FEM simulation (Supplementary Video S3). As predicted by the model, **Fig. 2 B, ii** shows that at the critical strain $\varepsilon_c \approx 1.42$, the initial concave boundary ribbon deforms into a straight line and remains straight upon further deformation, where we have $f\left(\overline{s}_b, \varepsilon\right) = \overline{s}_b$ and $g\left(\overline{s}_b, \varepsilon\right) = \sqrt{1.46 - (\varepsilon - 0.32)^2}$ with $\overline{s}_b \in [-1.32, 1.32]$ in the model (Supplementary Materials, Section S3). Correspondingly, the backbone profile (**Fig. 2 B, i**) transits from an initial sharp V shape to a smooth concave shape, which exhibits a sudden jump of the displacement along $z$-axis when the applied strain is slightly beyond $\varepsilon_c$. Further stretching results in the formation of the saddle shape with a concave backbone, which is consistent with both experiments (**Fig. 2 B, i-iii**) and FEM simulation results (**Fig. 2 B, iv**).

Based on the validated theoretical model, we further establish the general quantitative correlation between the boundary curvature $k_{bo}$ of 2D kirigami precursors and the Gaussian curvature $K$ of their popped 3D topologies at a given applied strain (see details in Supplementary Materials, Section S4). **Fig. 2 C** and **D** show the theoretically predicted 3D maps of the normalized Gaussian curvature $\overline{K}$ at the center point $C$ as a function of both normalized boundary curvature $\overline{k}_{bo}$ (see illustration of tuning different $k_{bo}$ in insets of **Fig. 2 C** and **D**) and applied strain $\overline{\varepsilon}$. It shows that for 2D kirigami precursors with either positive (**Fig. 2C**) or negative boundary curvature (**Fig. 2D**), generally, the absolute value of $\left| \overline{K} \right|$ increases with an increasing strain $\overline{\varepsilon}$ and $|\overline{k}_{bo}|$. Note that for the formed saddle shapes, we have $\overline{K} = 0$ before reaching the critical strain $\varepsilon_c$. At the onset of $\varepsilon_c$, $\overline{K}$ suddenly decreases due to a dramatic increase in the boundary curvature. Beyond $\varepsilon_c$, $\overline{K}$ barely changes because the boundary ribbon remains straight (**Fig. 2D**). Interestingly, **Fig. 2 E** shows that theoretically, the normalized variation of Gaussian curvature $\left| \Delta \overline{K} / \overline{K}_{\max} \right|$ increases approximately linearly with the normalized variation of boundary curvature $\left| \Delta \overline{k}_b / \overline{k}_{bo} \right|$ (slope $\approx 1$) (Supplementary Materials, Section S4), which is consistent with the experimental measurements. $\overline{K}_{\max}$ is the maximum Gaussian curvature at the center point. Specifically, this near-linear relationship holds regardless of the initial boundary curvature of a 2D kirigami precursor.

**Combinatorial designs for more complex 3D shapes**

Next, equipped with the knowledge of the correlation between the boundary curvature and the deformed 3D shapes, as well as the bistable feature (popping up or down in Fig. S1) in the discrete ribbons, we extend it to achieve more varieties of reconfigurable 3D shapes through combinatorial and tessellated designs.

We first explore the 3D shape shifting in 2D kirigami precursors through tessellating the three basic shapes in **Fig. 1**. **Fig. 3 *A, i*** shows a 2D diamond-shaped kirigami precursor composed of tessellated 2 × 2 square units with zero boundary curvature. Each unit has the same parallel cut pattern. Upon vertically stretching the 2D diamond precursor, both top and bottom square units pop up (represented by the symbol of "+" in the inset of **Fig. 3 *A, ii***) via out-of-plane buckling while the square units on two sides pop down (denoted by the symbol of "−"), generating a smiley 3D human face-like topology (**Fig. 3 *A, ii***). Notably, the eyes and mouth in the form of a hole are formed due to its discontinuous slope at the joints. The holes divide the face into eight independent popping regions (e.g., forehead, eyes, nose, cheek, mouth, and chin, etc.,). Thus, manipulating the bistable switch in each region, i.e., the popping directions of discrete ribbons, could generate more potential facial expressions. For example, flipping all the popping directions in the smiley face generates a sad face (**Fig. 3 *A, iii***), which can be reversibly switched. Localized flipping of two single ribbons in the eye area generates a face with eyeglasses (**Fig. 3 *A, iv***). Similarly, stretching an array of 3 × 1 rectangle units with identical parallel cuts and zero boundary curvature (**Fig. 3 *B, i***) leads to a sinusoidal wavy shape (**Fig. 3 *B, ii***), which can be reconfigured and reversibly switched to a coiled shape by flipping the popping direction in the central unit (**Fig. 3 *B, iii***). Furthermore, stretching the tessellated 2D precursor composed of two circular units bridged with a biconcave unit (**Fig. 3 *C, i***) generates an open Venus flytrap-like shape (**Fig. 3 *C, ii***). Notably, it can be rapidly closed by flipping the popping direction of the saddle-shaped junction to mimic the snapping of Venus flytrap (**Fig. 3 *C, iii***).

More complex 3D shapes can be generated by combining different boundary curvatures in stacked 2D kirigami precursors under uni-axial mechanical stretching, e.g., a 3D droplet-like shape (**Fig. 3 *D, ii***) and a vase-like shape (**Fig. 3 *E, ii***). We note that the smoothness of the backbone in the generated 3D shapes depends on the smoothness of the boundary in their corresponding 2D precursors (Supplementary Materials, Section S5). To generate the water droplet shape, we design a 2D precursor consisting of combined a straight line and a circular arc to mimic the 3D water droplet's backbone shape (**Fig. 3 *D, i***). Similarly, the vase-like shape (**Fig. 3 *E, ii***) is generated by stretching two stacked 2D precursors composed of a concave and convex boundary (**Fig. 3 *E, i***).

Furthermore, stretching multiple layers of similar semi-circular 2D precursors (**Fig. 3 *F*, *i***) generates a flower-like shape with multilayer pedals (**Fig. 3 *F*, *ii***).

**Inverse design strategy**

Existing methods of inverse design for target 3D curved shapes using the kirigami approach require complex algorithms to program heterogeneous local deformation among networked cut units (*33, 34*). Based on the information that discrete ribbons are geodesic curves of the deformed 3D shapes, we propose a straightforward inverse design strategy. It utilizes the geodesic curves extracted from the target shapes and the isometric mapping to prescribe the 2D precursors (**Fig. 4 *A***), which is, in principle, applicable to any target configuration.

To illustrate the strategy, we use the target shapes of a water droplet (**Fig. 4 *B***) and a vase (**Fig. 4 *D***) as two examples for the inverse design of the 2D kirigami precursors. As shown in **Fig. 4 *B*** and ***D***, we first approximate and represent the target shapes by deriving the shape functions of the backbone curve $B$ (highlighted in yellow color, see details in Supplementary Materials, Section S6), the geodesic curves $G$ (highlighted in orange color) approximated by elastica curves, and the boundary curve $\Gamma$ (highlighted in red color). Next, based on the isometric mapping from $G$ and $\Gamma$ in the target shape to the 2D precursor, we derive the shape function of the prescribed 2D boundary curve $\Gamma^P$ (Supplementary Materials, Section S6). The parametrization of $\Gamma$ and $\Gamma^P$ can be expressed in the form of $\tilde{r}_\Gamma = \left( x(s_b), y(s_b), 0 \right)$ and $\tilde{r}_{\Gamma^P} = \left( x^P(s_b), y^P(s_b), 0 \right)$, respectively, where the superscript $P$ represents the 2D precursor. Accordingly, the shape function of the boundary curve $\Gamma^P$ in the 2D precursor can be derived as (see details in Supplementary Materials, Section S6)

$$\begin{Bmatrix} x^P(s_b) \\ y^P(s_b) \end{Bmatrix} = \begin{bmatrix} \eta_x & 0 \\ 0 & \eta_y \end{bmatrix} \begin{Bmatrix} x(s_b) \\ y(s_b) \end{Bmatrix} \qquad [4]$$

where the parameters of $\eta_x$ and $\eta_y$ are related to the isometric mapping. Furthermore, the required strain $\varepsilon_{re}$ to form the target shape is given by $\varepsilon_{re} = \left[ x\left( \max(s_b) \right) - x^P\left( \max(s_b) \right) \right] \Big/ x^P\left( \max(s_b) \right)$ with $\max(s_b)$ being the maximum arc length of the boundary ribbon.

**Fig. 4 *C*** and ***E*** show the result of the inverse design of a water droplet and a vase after deploying the derived 2D kirigami precursors at an applied strain of $\varepsilon_{re} = 0.14$ and $0.07$, respectively. The inverse design result agrees well with the target shape denoted by the yellow curves. Notably, precise control of all the geodesic curves is not necessary for our proposed inverse design approach since it only needs the information of one representative geodesic curve and one boundary curve in the target surface. Thus, such a strategy could significantly simplify the calculation (Supplementary Materials, Section S6). The accuracy can be further improved by increasing the number of geodesic curves deriving from the target surface and optimizing the approximation of geodesics.

**Remote magnetic actuation and proof-of-concept potential robotic application**

In addition to the contact-based mechanical actuation for shape morphing, we also demonstrate the capability of achieving similar 2D-to-3D shape shifting and bistability in kirigami sheets through remote magnetic actuation. By attaching a magnetic pad to one end of the 2D circular kirigami precursor with the other end fixed, **Fig. 5 *A*** shows a similar shape shifting into a spheroidal shape under an applied remote translational magnetic field to that under mechanical stretching (Supplementary Video S4). Given the flexibility of the discrete ribbons, the 2D circular kirigmi precursor could also be twisted into a pine-cone shape under either an applied rotational magnetic field or a mechanical twisting (Supplementary Video S5 and Fig. S9a). Similarly, by attaching magnetic thin polymeric stripes to the discrete ribbons (**Fig. 5 *B*, *i***), we could use a remote translational magnetic field to fast switch the bistable states in the ribbons to reconfigure the kirigami structure. **Fig. 5 *B*, *ii-v*** show a sequential snapping of the discrete ribbons in the spheroidal structure (Supplementary Video S6), where its sequence could be tuned via the distribution of the magnetic polymers in the 2D kirigami precursor (Supplementary Materials, Section S7).

Lastly, to demonstrate the advantages of both remote magnetic actuation and the proposed kirigami design, we explore its potential application in designing an untethered predator-like flexible robot as a proof of concept. The flexible kirigami robot mimics the rolled-up pill millipede to predate a target object, where a tiny rock is used to simulate the prey and its irregular shape is to demonstrate the capability of delicate manipulation and adaptability of the kirigami flexible body (**Fig. 5 *C*, *i***). As shown in **Fig. 5 *C*, *ii-v*** and Supplementary Video S7, the predating and releasing process consists of three steps. (1) Catching by moving and rotating itself around *x*-axis to cover the object under an applied rotational magnetic field around *x*-axis in **Fig. 5 *C, ii***, followed by relocating the object to a different spot under a translational magnetic field (**Fig. 5 *C, iii***). (2) "Swallowing" the object through curling its flexible body along the minor axis to wrap the irregular rock inside under a rotational magnetic field around *y*-axis (**Fig. 5 *C, iv***). It should note that the flexible body adapts and conforms to the irregular shape of the rock by tightly wrapping around it, the deformation and dynamic shape morphing of which is challenging to be realized by contact-based mechanical actuation. (3) Carrying the "swallowed" object to a new spot followed by object releasing through recovering to its undeformed configuration (**Fig. 5 *C, v***). An alternative way to "swallow" the object is also demonstrated by curling its body along the major axis under a rotational magnetic field around *x* axis (Fig. S9b and Supplementary Video S8), which mimics the snake-coiling. The discrete ribbon-based kirigami design shows the unique features of flexibility, conformability, and adaptivity, which can help to accommodate remotely actuated complex dynamic shape morphing that could not be achieved by mechanical actuation, and thus facilitate the target robotic multitasks of locomotion and manipulation during predation.

## DISCUSSION

We proposed a new way of utilizing the cut boundary curvature to guide the formation of controllable and reconfigurable complex 3D curved shapes in kirigami sheets patterned with simple parallel cuts. Such a strategy is validated through combined theoretical modeling, FEM simulations, and experiments. The unique feature of discrete cut ribbons as geodesic curves of the deformed 3D shapes largely simplifies the inverse design. Beyond contact-based mechanical stretching, we demonstrated that the strategy could also apply to the remote magnetic actuation for enhanced shape-morphing capability in kirigami sheets imparted with potential untethered robotic functionality.

It should note that the shape shifting in the studied thermoplastic kirigami sheets is fully reversible, i.e., the generated 3D shape will return to its original flat form after the external actuation is removed due to elastic deformation in the kirigami structure. To preserve the deformed 3D shapes, we could utilize the shape memory properties of the PET polymer upon heating above its glass transition temperature (*39*). We use thermal treatment under 120 ºC to treat the 3D shapes held at an applied stretching strain for a period of 120 min and cooled down to the room temperature to fix the deformed configuration (see the demonstration of a fixed 3D spheroidal shape in Fig. S10). Notably, the preserved 3D configuration can be further deformed and recover to its 2D flat precursor shape upon another thermal treatment (Fig. S10).

Despite the demonstration of programmable shaft shifting in the thermoplastic kirigami sheets, we envision that the proposed strategy is material and scale independent. We note that despite the large applied stretching strain $\varepsilon$, the maximum principal strain $\varepsilon_{max}$ in the buckled ribbons with thickness of 127 μm remains small ($\varepsilon_{max} < 1\%$ for $\varepsilon > 50\%$, Fig. S11, Supplementary Materials, Section S3), e.g., $\varepsilon_{max} = 0.4\%$ in the deformed spherical shape at $\varepsilon = 30\%$ (**Fig. 2 *A*, *iv***) and $\varepsilon_{max} = 0.6\%$ in the saddle shape at $\varepsilon = 147\%$ (**Fig. 2 *B*, *iv***). Considering the small peak tensile strain in the buckled ribbons and its linear relationship with sheet thickness $t$, i.e., $\varepsilon_{max}$ decreases linearly with $t$, we envision that the proposed kirigami strategy could also be applied to design shape-morphing and stretchable structures and devices made of other functional materials such as metals and even semiconductors at small scales, as well as other stimuli-responsive materials actuated by temperature, electrical, and magnetic field, etc. This work could find potential applications in designing stretchable electronics, reconfigurable devices, soft robots, and mechanical self-assembly fabrication.

## MATERIALS AND METHODS

Details on fabrication of 2D kirigami precursors, finite element simulation, theoretical modeling, and inverse designs are described in the Supplementary Materials.

**Acknowledgments: Funding:** J. Y. acknowledges the funding support from the National Science Foundation under award number CAREER-2005374 and 2013993. **Author contributions:** Y.H. and J. Y. designed research; Y. H., Y. C., and Y. Li performed research; Y. H., Y. C., Y. Li, Y. Z., and J. Y. analyzed data; Y. H. and J. Y. wrote the paper.

**Competing interests:** The authors declare no competing interests. **Data and materials availability:** All data is available in the main text or the supplementary materials.

## SUPPLEMENTARY MATERIALS

Text

Fig. S1 to S11

Movie S1 to S8

Reference Notes

**Figures:**

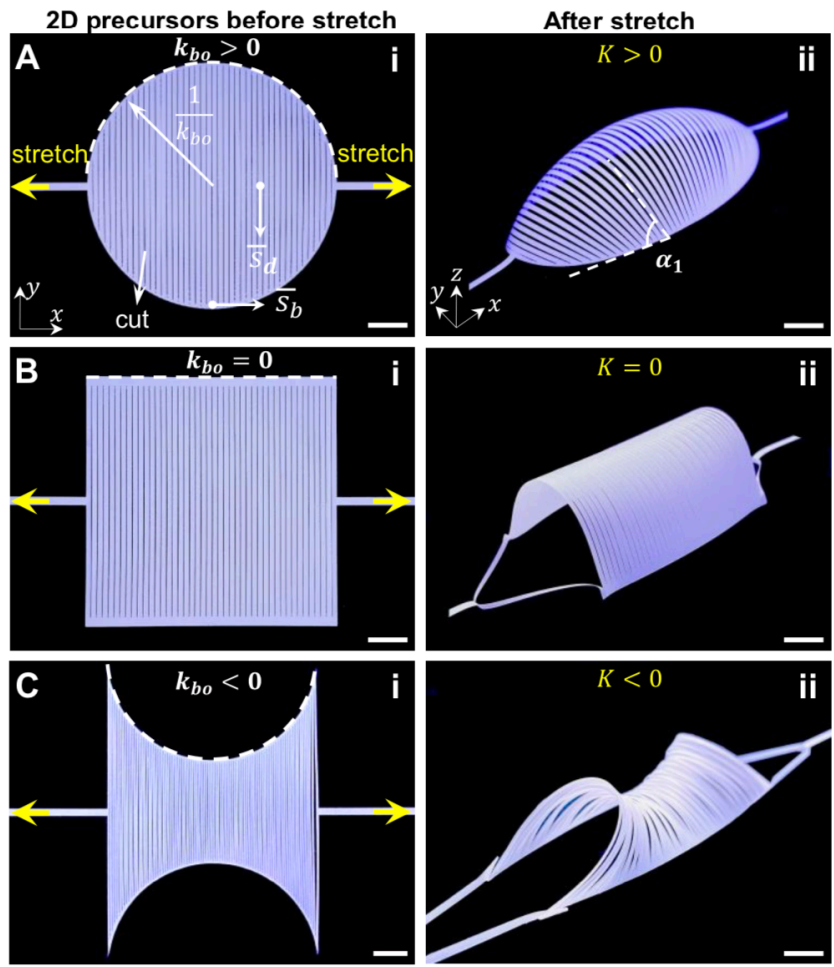

**Fig. 1 Shape shifting of 3D curved topologies from 2D kirigami sheets with different cut boundary curvatures subject to uniaxial tension.** (*A-C*) Left column: 2D precursors of three kirigami sheets patterned with parallel cuts but different boundary curvatures $k_b$ highlighted in dashed white curves. Circular (*A*), square (*B*), biconcave (*C*) samples with positive, zero, and negative boundary curvature, respectively. Right column: Formed 3D curved shapes with different Gaussian curvature $K$. (*A*) Spheroidal shape with $K > 0$ at an applied strain of 0.30. (*B*) Cylindrical shape with $K = 0$ at an applied strain of 0.65. (*C*) Saddle shape with $K < 0$ at an applied strain of 1.47. Scale bars = 10 mm.

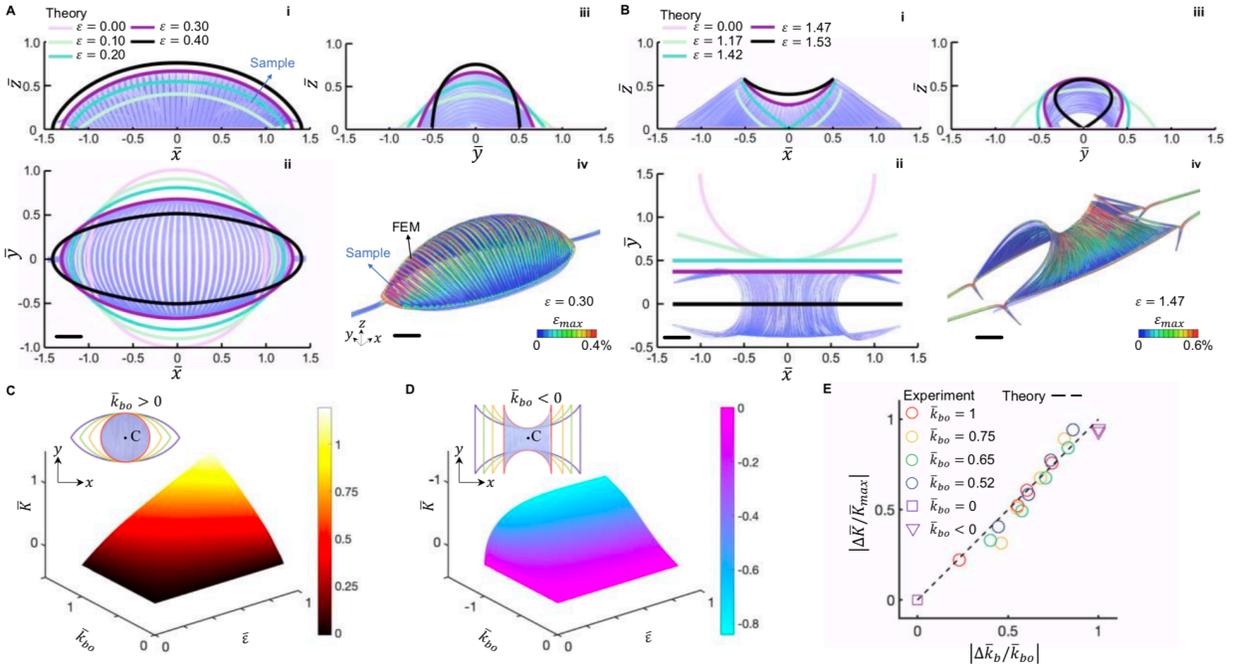

**Fig. 2 Quantifying the 3D shape shifting through analytical modeling and simulation.** (***A***, ***B***) Predicted shape changes with the applied strain $\varepsilon$ in the samples of spheroidal (***A***) and saddle shapes (***B***). (***i***) is the front-view profile. (***ii***) is the top-view profile. (***iii***) is the side-view profile. (***iv***) is the overlapping of FEM simulation results (contours of the maximum principal strain $\varepsilon_{max}$) with the experimental image at $\varepsilon = 0.30$ (***A***) and 1.47 (***B***). (***C***, ***D***) Theoretically predicted 3D maps of the normalized Gaussian curvature $\overline{K}$ at the center point $C$ as a function of the normalized boundary curvature $\overline{k}_{bo}$ in 2D kirigami precursors (insets) and the applied strain $\varepsilon$ for the cases of spherical (***C***) and saddle (***D***) shapes. (***E***) Theoretical and experimental results of the approximately linear relationship between the normalized variation of the Gaussian curvature $\left| \Delta \overline{K} / \overline{K}_{\max} \right|$ and the normalized variation of the boundary curvature $\left| \Delta \overline{k}_b / \overline{k}_{bo} \right|$. Scale bars = 10 mm.

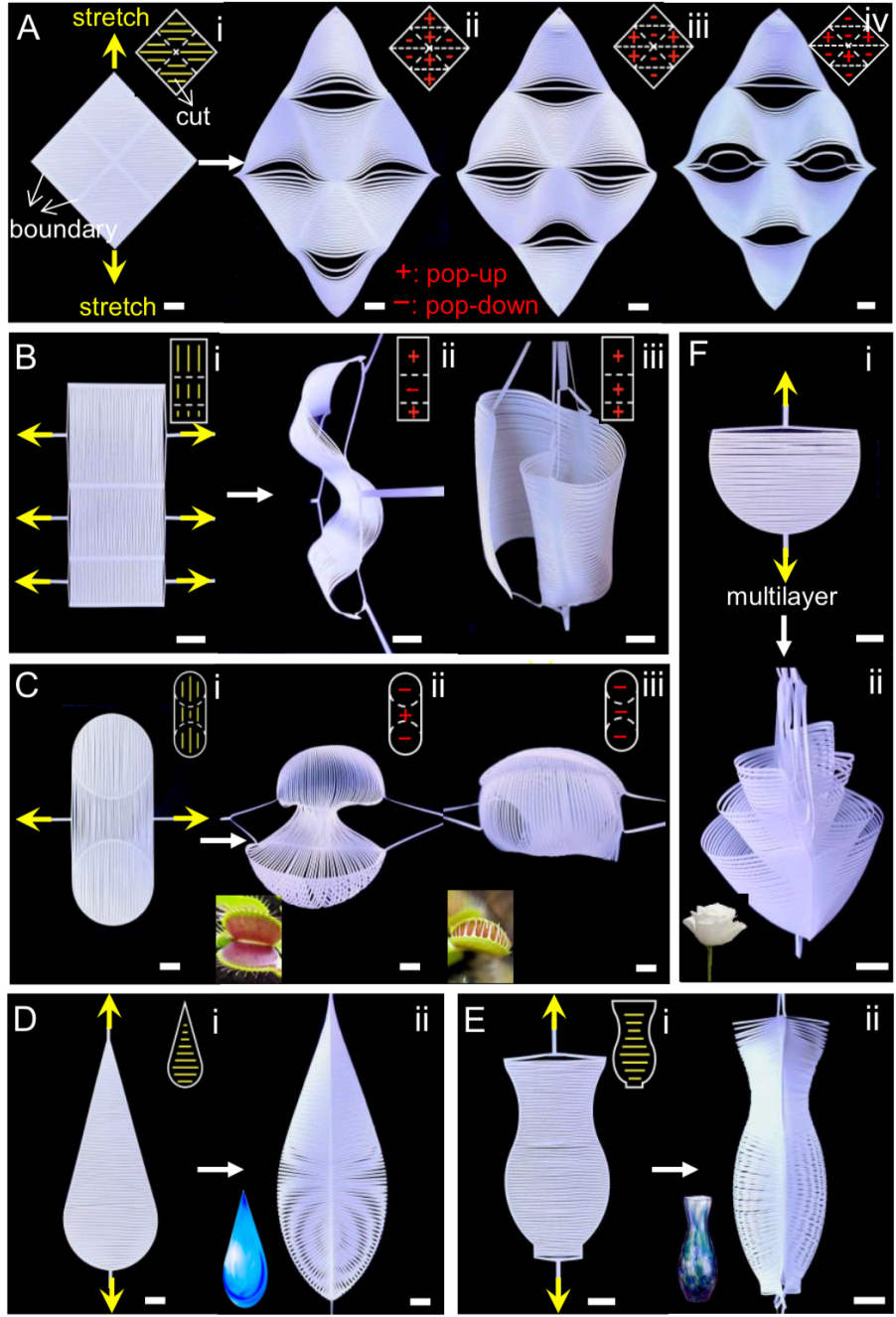

**Fig. 3 Combinatorial designs of 2D kirigami precursors for complex 3D shapes under uniaxial tension.** (*A-C*) Reconfigurable 3D shapes through bistability of discrete ribbons. (*A*, *i*) 2D precursor composed of 2 × 2 square units with zero boundary curvature. (*A*, *ii-iv*) uni-axial stretching induced reconfigurable human face-like topologies with switchable smiley (*ii*) and sad (*iii*) expressions, as well as eyeglasses (*iv*) by tuning the popping directions of ribbons (insets). (***B***,

*i*) 2D precursor composed of 3 × 1 rectangle units with zero boundary curvature. (***B***, ***ii-iii***) Formation of switchable sinusoidal wavy and coiled shapes. (***C***, ***i***) 2D precursor composed of an array of two circular units bridged with a biconcave unit. (***C***, ***ii-iii***) Formation of switchable opening and closing of Venus flytrap-like shapes. (***D-E***) Formation of a 3D droplet-like shape (***D***, ***ii***) and vase-like shape (***E***, ***ii***) by uni-axially stretching 2D precursors with different combined boundary curvatures (***i***). Insets show the image of a droplet and a vase. (***F***) Formation of a flower-like shape (***ii***) by uni-axially stretching multiple layers of semi-circular 2D precursors (***i***). Scale bars =10 mm.

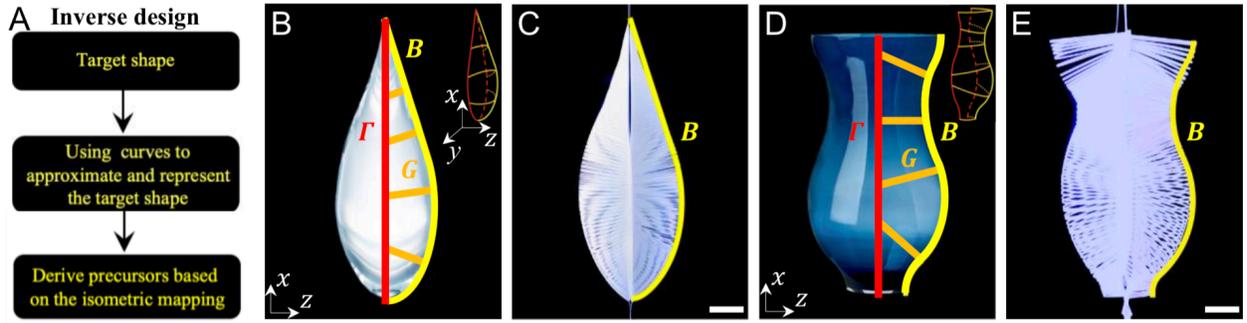

**Fig. 4 Inverse design of 3D shapes.** (***A***) Flow diagram of the inverse design. (***B***) Schematic of using curves to approximate and represent the target shape (side view of a waterdrop). The inset shows an isometric view. Red, orange, and yellow curves are the boundary curve $\Gamma$, geodesic curve $G$, and backbone curve $B$, respectively. (***C***) Experimental inverse-design result of the waterdrop shape formed by a 2D kirigami precursor subject to uniaxial tension. The yellow curve is the backbone in the target shape. (***D***) Schematic of using curves to approximate and represent the target shape (side view of a vase), with an isometric view showing in the inset. (***E***) Experimental inverse-design result of the vase shape formed by a 2D precursor subject to uniaxial tension. Scale bars = 10 mm.

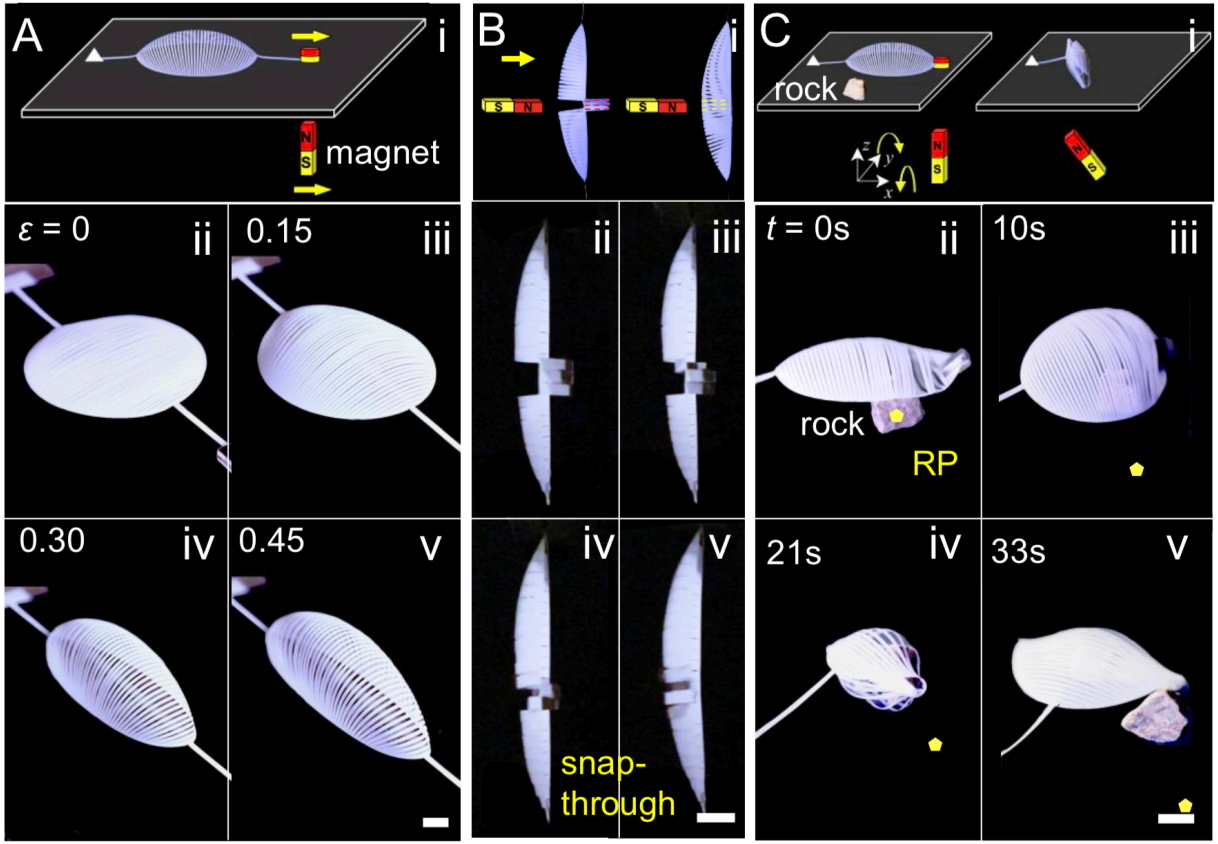

**Fig. 5 Remote magnetic actuation of dynamic shape shifting.** (*A-C*, *i*) Schematic illustration of remote magnetic actuation. (*A*, *ii-iv*) An actuated spheroidal shape under translational magnetic field. (*B*, *ii-iv*) A moving magnetic field triggers the sequential snapping of the bistable central ribbons with magnetic polymers. (*C*, *ii-iv*) Proof-of-concept demonstration of an untethered predator-like flexible kirigami robot under rotating magnetic field to mimic rolled-up pill millipedes for predating, relocating, and releasing target object (a small irregular rock). RP represents the reference point. Scale bars = 10 mm.